# ANALYSIS OF SOME EXACT PROPERTIES OF THE NONLINEAR EQUATION OF THE SHRÖDINGER DESCRIBING THE PROCESS OF FILAMENTATION

A.D.Bulygin

Abstract:: The properties of the integral of motion and the evolution of the effective radius of the light beam are analyzed in the framework of the stationary model of the nonlinear Schrödinger equation describing filamentation. Within the framework of such a model, it is shown that filamentation is limited only by dissipative mechanisms.

## 1. Introduction

The mathematical model describing the phenomenon of self-focusing and filamentation is a nonlinear Schrödinger equation (NSE). In the stationary version, which we will consider in this paper, it has the dimension (2 + 1). It should be noted that, in contrast to (NSE) with a dimension (1 + 1), for which the theory of the inverse scattering problem (ISP) was constructed [1, 2], where, accordingly, exact soliton solutions for NSE with cubical local nonlinearity in higher dimensions were obtained, analogous exact solutions are unknown. Nevertheless, recently various approximate and asymptotic methods for studying the NSE equations are actively developing, including nonlocal nonlinearity, see, for example, [3,4]. However, for NSE with local cubic nonlinearity in higher dimensions, a very limited set of precise analytical results is known. This property of the global collapse in the Kerr medium [1,5], the Townes solution, which, incidentally, is also not analytical [6], but in accordance with the theory of Bespalov-Talanov instability [5] it is also unstable.

It turns out, as we shall show below, that for a more general case of local nonlinearity of the type [7], considered, for example, in works on filamentation [8,9], one can obtain a number of simple and useful relations for integral parameters. Surprisingly, these, in general, trivial relationships in the literature on self-focusing and filamentation have not yet been given due attention, see, for example, reviews. [5,10]. The reason is that, in connection with the development of modern high-performance numerical methods, the problem of filamentation is investigated mainly on the basis of numerical experiments [5, 10]. In addition, as noted above, approximate analytical methods for investigating this problem also develop, such as the variational method [9, 11] and even renormalization group analysis methods [12].

However, these approximate methods use either the approximation of trial functions [9, 11] or the eikonal approximation [12] (this approximation, as shown in our papers [13, 14], is not correct in the filamentation process). In our opinion, the choice of localized functions in the form of Gaussian or similar ones is not physical, since it does not allow describing either the formation of the annular structure characteristic for filamentation or the growth of the integral size of the light beam, discovered in [15]. As for direct numerical simulation, it is usually performed blindly, i.e. without realizing the test for exact model solutions, because of their absence. In the case of a stationary formulation of the problem, when the integrals of motion are known [7], as shown in our recent paper [16], such a "blind" approach (realized by the standard numerical schemes for the study of filamentation) leads to a substantial distortion of the results for some important physical quantities .

Perhaps, one of the few papers about filamentation on which it is worth paying attention is the paper [17] in which, albeit approximately and for a limited class of beams, a result is obtained for a sufficient condition for self-channeling. In our short note, we generalize this result (if we mean an infinite process for self-channeling) for an arbitrary class of sheaves and at the same time without any approximations, that is, we obtain it exactly from a direct analysis of the integral properties of a light beam and known integrals of motion.

The presence of exact results in the context of the above situation (in such a field of research as filamentation) is of undoubted interest, even if we consider simple stationary models of filamentation.

Also, to illustrate our exact results, based on our well-developed numerical scheme for the solution of NSE, [16], we investigate the dynamics of quantities (which are integrals of motion for the conservative case) for the case of inclusion of nonlinear absorption mechanisms.

## 2. Stationary conservative model of filamentation

Here we will consider one of the simplest versions of NSE, considered, for example, in [5,8].
This NSE variant corresponds to the conservativeness and stationarity approximation in the description of filamentation. In these approximations, the NSE for the complex envelope of the light field $U(\vec{r},z)$ in dimensionless variables has the form:

$$-2i\frac{\partial}{\partial z}U + (n_{ker}I + n_{sat}I^{m-1} + \Delta_\perp)U = 0 \qquad (1)$$

Here $\vec{r}=(x,y)$ - transverse coordinates, $\Delta_\perp = \partial_x^2 + \partial_y^2$ ; $I = UU^*$; $n_{ker}, n_{sat}, m$ - model parameters that take into account the effects of cubic self-focusing $n_{ker} < 0$ in a nonlinear medium and defocusing ($n_{sat} > 0$), due to the inclusion of higher nonlinearities of the order $m$. It should be noted that the higher nonlinearities can be simulated either by the plasma defocusing [18] arising in the medium, or by the saturation effect of the Kerr (cubic) nonlinearity [8,9,12,17]. This equation is a particular case of a general form of equations of the type of the nonlinear Schrödinger equation, considered, for example, in [6].

In this case, the dynamic equation is a variational one, resulting from the action:

$$S = \int 2i(\partial_z U^* U - \partial_z U U^*) - \frac{1}{4}(2|\nabla_\perp U|^2 - \Delta_\perp U U^* - U\Delta_\perp U^*) + n_{ker}I^2/2 + n_{sat}I^m/m)d\vec{r}dz \qquad (2)$$

Equations of motion can also be written in the Hamiltonian formulation [1]:
$$-2i\partial_z U - \delta_{U^*}H = 0.$$

Where the Hamiltonian function is defined by the expression:
$$H \equiv \int h d\vec{r} = H_d + H_{kerr} + H_{sat}$$

Here:
$$H_d \equiv \int h_d d\vec{r} = \frac{1}{4}\int (2|\nabla_\perp U|^2 - \Delta_\perp U U^* - U\Delta_\perp U^*)d\vec{r}$$

$$H_{ker} \equiv \int h_{ker} d\vec{r} = \int \frac{n_{ker}I^2}{2}d\vec{r}$$

$$H_{sat} \equiv \int h_{sat} d\vec{r} = \int \frac{n_{sat}I^m}{m}d\vec{r}$$

Conservation laws that correspond to two global NSE symmetries (the outer one is the shear transformation over the evolutionary coordinate $z \to z+a$ and internal - the transformation of the phase shift of the complex field $U \to Ue^{ia}$), can be written in the form of continuity equations:

$$\partial_z I + \nabla_\perp \mathbf{P}_\perp = 0 \Leftrightarrow \partial_z \int I d\vec{r} = 0 \qquad (3a)$$
$$\partial_z h + \nabla_\perp \mathbf{Q}_\perp = 0 \Leftrightarrow \partial_z \int h d\vec{r} = 0 \qquad (3b)$$

Where $\mathbf{P}_\perp$ and $\mathbf{Q}_\perp$ the corresponding current vectors, determined directly from (1). So the magnitude $E = \int I d\vec{r}$ corresponds in the terminology of optics to the normalized beam power or in terms of quantum mechanics to the number of particles, and $H = \int h d\vec{r}$ in the terminology of the optics of nonlinear divergence or energy in terms of quantum mechanics. In particular, the expression for the Umov-Poyntyg vector has the form:

$$\mathbf{P}_\perp = \frac{U^*\nabla_\perp U - U\nabla_\perp U^*}{2i}$$

The equation for this vector can be obtained from (1) after a series of elementary transformations and it has the form:

$$\partial_z \mathbf{P}_\perp = \nabla_\perp (h_g + \frac{m-1}{m}h_{sat})/2 \qquad (4)$$

Where $h_g \equiv h_d + h_{\text{ker}}$. From the relations (3-4) for the square of the effective radius of the beam, defined as the second spatial moment of the intensity of the light field $R_e^2 = \int I \vec{r}^2 d\vec{r}$, follows the relation[1]:

$$\frac{\partial^2}{\partial z^2} R_e^2 = 2(H_g + 2\frac{m-1}{m} H_{sat}) \qquad (5)$$

Where we introduced the notation $H_g \equiv \int h_g d\vec{r} = H_d + H_{ker}$. This relation, as is well known, means that in a purely Kerr medium (m = 2) with a negative value of the invariant (standing in this case on the right-hand side), a sufficient collapse condition is realized [2.5].

Further, note that usually the parameter m in (1) has a sufficiently large value (for water it is equal to 4-5, and for air 8-10) in order that the laser radiation propagation region can be divided with a physical accuracy into regions ( sections in $z$ ) , where there is filamentation $H_{sat} > 0$ (i.e., there is a plasma formation or higher mechanisms of nonlinearity appear, due to the saturation of the Kerr nonlinearity) and where it is not present $H_{sat} = 0$. This partition corresponds to the partition of space into regions where the maximum intensity along the profile is close to a certain value typical for filamentation [5, 10]. Then the last assertion can be formulated from the last relation.

**Proposition: A prerequisite for the possible termination of filamentation is the condition $H_g \geq 0$**

**Proof:**

Indeed, suppose that $H_g < 0$. Let at the beginning of the track filamentation has not yet begun, i.e. $H_{sat} = 0$, then the bundle as a whole collapses in accordance with the virial theorem in a Kerr medium. Further, when approaching the collapse point, higher nonlinearity mechanisms are included (if the higher nonlinearities model the plasma, then this corresponds to "plasma formation"), i.e. $H_{sat} > 0$ and it stops the collapse [5,8,9,10]. Suppose that somewhere in the filamentation, still ended, that is, the condition $H_{sat} = 0$, but due to the fact that, as before, we have a relation for the total integral of motion $H_g < 0$, hence again it follows that the bundle as a whole will again start to collapse, and hence the filamentation will never finally be completed.

Formally, this statement can be written in the implicative form $(L_f < \infty) \Rightarrow (H_g \geq 0)$, где $L_f$ - the filamentation length is defined as the distance between the extreme points where the inclusion of terms with higher nonlinearity is realized or in other words the intensity reaches the filamentation level. Here from condition $H_g \geq 0$ nothing, generally speaking, does not follow in the local sense. This necessary condition for the limited length of filamentation can be conveniently read in reverse order, namely $(H_g < 0) \Rightarrow (L_f \rightarrow \infty)$ , i.e. condition $(H_g < 0)$ is sufficient to ensure that the length of the filamentation is infinite.

Thus, a sufficient condition for the formation of infinite filamentation or, in other words, an attractor, is established if we understand a joint system of equations for the field and the medium as a dynamical system. The conservative model considered by us, although it is used in the investigation of MFLI in air and water, etc., however, is a rough idealization with the loss of the physics of the process. Nevertheless, the use of this model is possible for some media, where the threshold for the manifestation of the mechanisms of saturation of the cubic nonlinearity is the smallest in comparison with the thresholds for other higher nonlinear mechanisms accompanied by dissipation, and in this case indeed, when the threshold radiation powers are reached, a very long region filamentation.

In the next section, we will consider a model with dissipation and on the basis of numerical calculation we illustrate, including the conclusions made in this section.

**The model with dissipation**

Let us consider a stationary model with dissipation see [18]. In this case, the NSE will take the form:

---

[1] This relation is sometimes called the virial theorem [2], while in [2] an incorrect expression is given on the right-hand side of the equation for the effective radius.

$$-2i\frac{\partial}{\partial z}U + (n_{ker}I + n_{sat}I^{m-1} + i\alpha_N(I) + \Delta_\perp)U = 0 \tag{6}$$

Where the coefficients of nonlinear absorption $\alpha_N$ are associated with the processes of multiphoton photoionization of the medium $\alpha_N = I^{K-1}L_R/L_{MPA}$; $L_R = k_0R_0^2/2$ – Rayleigh beam length; $k_0 = 2\pi/\lambda_0$ – wave number; $\lambda_0$ – carrier wavelength; $R_0$ – initial mean-square radius of the laser beam $L_{MPA}$ - length $K$-photon absorption (in our case $K=8$). In this case, the integral identities (3) are transformed to the form:

$$\partial_z \int I d\vec{r} + \int \alpha_N I d\vec{r} = 0 \tag{7a}$$

$$\partial_z \int h d\vec{r} + \int \alpha_N(I)(h_d - \Delta_\perp I + 2n_{ker}I^2 + mn_{sat}I^m)d\vec{r} = 0 \tag{7b}$$

The additional term due to dissipation in relation (7b), unlike (7a), is not sign-definite and is determined by the form of the dynamic field. Accordingly, the behavior of the function $H_g$ for each specific implementation requires a special study. Than we are below and we will deal with an example of a number of problems in a radial formulation.

It should be noted that for the first time the correct numerical scheme for the solution of NSE with the correctly chosen adaptive grid construction algorithm ensuring the fulfillment of conservation laws (2) was developed by us in [16]. It is on the basis of this numerical scheme that we will investigate the behavior of the quantities $H_g$ and $E$ in the process of filamentation. The initial beam profile was set in Gaussian and super-Gaussian shapes, the beam radius was chosen to be 1 mm and 2 mm, the power varied from 3 to 9 $P_{cr}$ ($P_{cr}$ =3.2 GW for air), considered collimated and focused ($F = L_r$) beam. Below is an example of the results of calculations for quantities $H_g, E$.

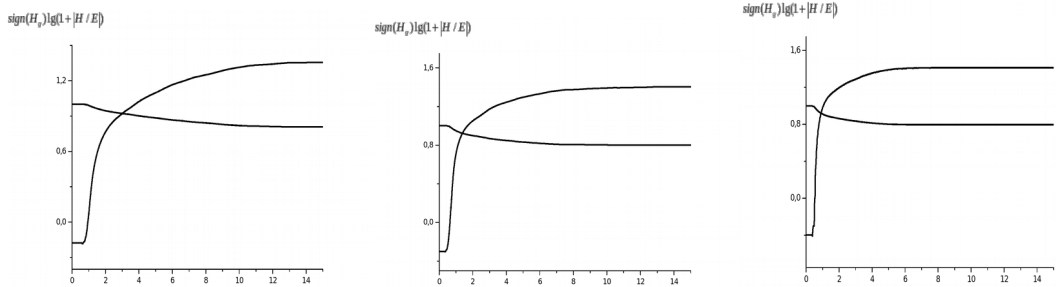

Fig.1 Dependence on the beam power normalized to its initial value $E/E_0$ along the propagation path and, respectively, $sign(H_g)\lg(1+|H/E|)$ for different values of initial power

From these graphs it can be seen that the function $H_g$ monotonously increases by saturation. Constancy along the trace function $H_g$ and $E$ corresponds to the absence of filamentation, so absorption is included in the same place where higher refractive nonlinearities are included. It is also clear that $H_g$ passes the zero value much earlier than the filamentation ends, which again confirms our statement that the condition $H_g < 0$ is a sufficient, but necessary condition for filamentation. Note that this behavior of the function $H_g$ was universal in all our numerical experiments.

Also, based on these graphs, it can be assumed that between the change $H_g$ and $E$ there is some simple connection. For this assumption, we have $\Delta \bar{E}$ on the values $\Delta E_{pf}/\Delta H_{g(pf)}$ (pf-post filamentation) and deduced the corresponding curves in Fig. 2.

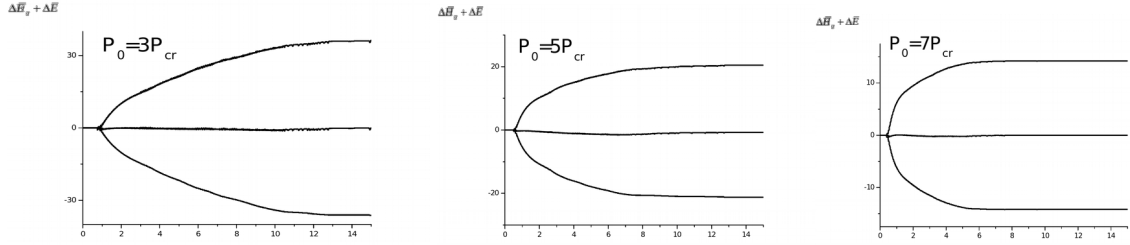

Fig. 2. Dependence on the normalized quantities $\Delta \bar{E}$ and $\Delta \bar{H}_g$ and $\Delta \bar{H}_g + \Delta \bar{E}$, to the corresponding post-filational values, from the propagation distance for different initial power values.

From these graphs it can be seen that the quantity $\Delta E$ with high accuracy is proportional to $\Delta \bar{H}_g$ i.e. the following relation holds:

$$H_g(z) = H_g(0) + \gamma(\Delta E) \tag{8}$$

Where $\gamma$ a constant along the propagation path characteristic for each numerical experiment. Our fairly extensive numerical study has shown that in each numerical experiment $\gamma$ along the route remained constant with great accuracy. Wherein $\gamma$ essentially depends on the initial conditions (focal length, power, radius), varying by a factor of several times in our numerical experiments. This fact seems somewhat surprising to us. Indeed, the constancy along the trace of magnitude $\gamma$ could be explained by the weak dependence of the field characteristics in the filamentation region on the initial conditions [5], but a strong dependence $\gamma$ from the focal length, power and initial radius do not support this explanation. From the totality of the listed facts with sufficient degree of credibility, we can conclude that the quantity $\gamma$ is a certain integral of the motion for the NSE in the form (6) (in another equivalent formulation, the integral is the quantity $H_g(z) - \gamma E(z)$ ). However, the strong evidence of this hypothesis and the explicit expression $\gamma$ from $U_0$.

**Conclusions**

Thus, in this paper, in the framework of the stationary conservative model of filamentation, a sufficient condition for unrestricted filamentation is established rigorously (without any approximations) on the basis of analysis of the integrals of motion: $H_g < 0$. A model with a nonlinear dissipation mechanism is also considered. On the basis of a numerical experiment, it is established that, due to dissipation mechanisms, an increase in the value $H_g$, which ensures the output of the system from the filamentation mode. In addition, it has been established that with a high degree of accuracy in the process of filamentation (with dissipation), the increment in the value $H_g$ in proportion to the change in $E$, the coefficient of proportionality $\gamma$ depends only on the initial characteristics of the field and can be regarded as the integral of motion for NSE with dissipation.

How the conclusions made in this paper are correlated with the complete nonstationary model of filamentation requires a separate study. To do this, we first need to write the integrals of motion for the nonstationary model of NSE (which at the moment is not available), secondly, the development and implementation of correct numerical schemes for solving nonstationary NSE (which are also not available at the moment). It is the solution of these tasks that we plan to devote to our next work.

«The research was carried out with the financial support of the Russian Foundation for Basic Research in the framework of the scientific project No. 14-28-02023 of the office of the Russian Research Foundation (Agreement 15-17-10001).»